\documentclass{kapproc}
\usepackage{graphicx}

\def\etal{{\em et al.}}
\def\deg{^\circ}

\kluwerbib

\begin{document}
\articletitle{Relativistic Effects on the Appearance of a Clothed Black Hole}
\author{Xiaoling Zhang, S.Nan Zhang, Yuxin Feng, Yangsen Yao}
\affil{Physics Department, University of Alabama in Huntsville, 
Huntsville, AL, USA\\
National Space Science and Technology Center, Huntsville, AL, USA}
\email{zhangx@email.uah.edu}
\begin{abstract}
For an accretion disk around a black hole, the strong relativistic effects 
affect every aspect of the radiation from the disk, including its spectrum, 
light-curve, and image. This work investigates in detail how the 
images will be distorted, and what the observer will see from different
viewing angles and in different energy bands.
\end{abstract}
\section*{Introduction}

Black holes are believed to be very common in the universe. 
A black hole itself, as indicated by its name, is invisible. 
But the interactions between a black hole and other objects can generate 
spectacular phenomena and make the hole ``visible''. 

A ``clothed black hole'', a black hole with an accretion disk
(Luminet, 1978), is one of the most important laboratories to study
relativity. Strong relativistic modifications on energy spectra,
light curves, power spectra etc., have been observed with the many detectors
on the ground and in space, yet we have never observed 
any black hole image due to the limited spatial resolution of 
currently available instruments.
People have been curious about what a black hole or a clothed black hole 
would look like. 
Luminet (Luminet 1978) studied 
the Schwarzschild black hole case and gave a simulated photograph
of a spherical black hole with a thin accretion disk. 
Fanton \etal (1997) developed a semi-analytical approach 
and a code of calculating photon trajectories in Kerr metric, 
calculated the redshift maps of both direct and secondary images, 
and the temperature map.
We used their code in the calculation of the
disk images in this work. 

The observable disk image is actually the flux map. 
However, the total flux (in the whole energy range) map might not be 
a proper object, because each instrument has its sensitive energy range;
it can only detect photons in this range, thus instruments sensitive to
different energy ranges will see different pictures.

In section~\ref{section:redshiftmaps} we demonstrate 
how the relativistic effects in a Kerr space will work on various elements
of a black hole system. In section~\ref{section:images} 
we give the images in different energy ranges, followed by
summary and discussion.

\section{Red-shift maps of disks}
\label{section:redshiftmaps}

\begin{figure}[ht]
	\centering
        \begin{minipage}{.48\textwidth}
                \centering
		\includegraphics[width=\textwidth]{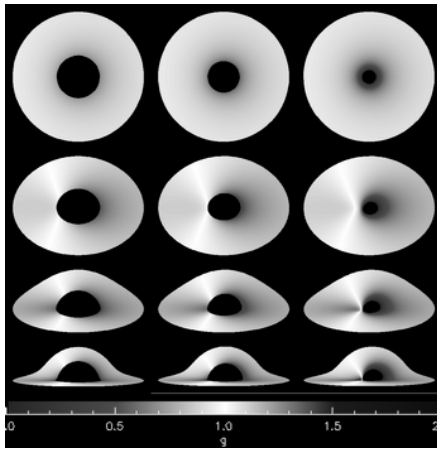}
        \end{minipage}
        \begin{minipage}{.5\textwidth}
                \centering
		\caption{
Red-shift map and overall shape of accretion disks around black holes. 
$g$ is the ratio between the observed energy of a photon and the energy
in the rest frame of the disk.
Top to bottom: $\theta=5\deg, 45\deg, 70\deg, 85\deg$,
$\theta$ is the inclination angle 
(the angle between the normal of the disk and the line of sight).
Left to right: $a=0.01, 0.5, 0.998$, 
where $a$ is the specific angular momentum of the black hole.
The white zones stand for the regions with zero red-shift 
(following Fanton \etal (1997) ).
Left-hand side of the disk is approaching the observer and blue-shifted.
}
        \end{minipage}
\label{fig:redshiftmap}
\end{figure}

\begin{figure}[ht]
	\centering
        \begin{minipage}[t]{.48\textwidth}
                \includegraphics[width=\textwidth]{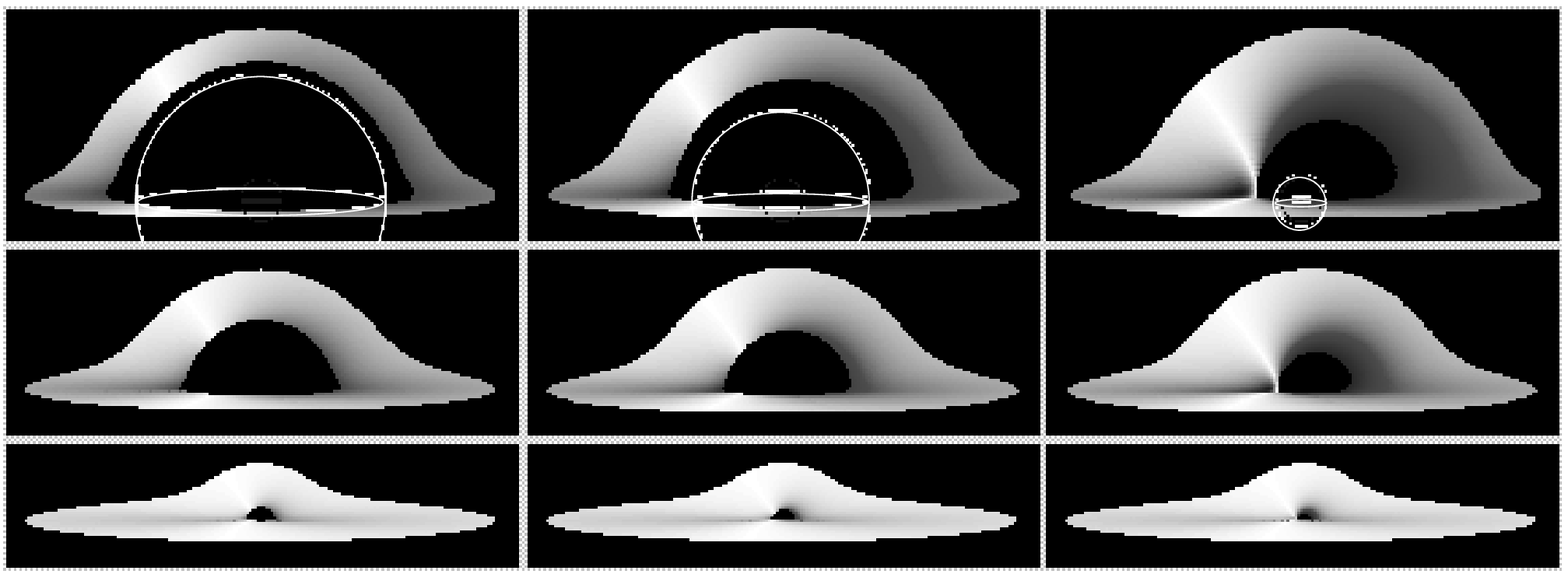}
        \end{minipage}
        \begin{minipage}[t]{.48\textwidth}
                \includegraphics[width=\textwidth]{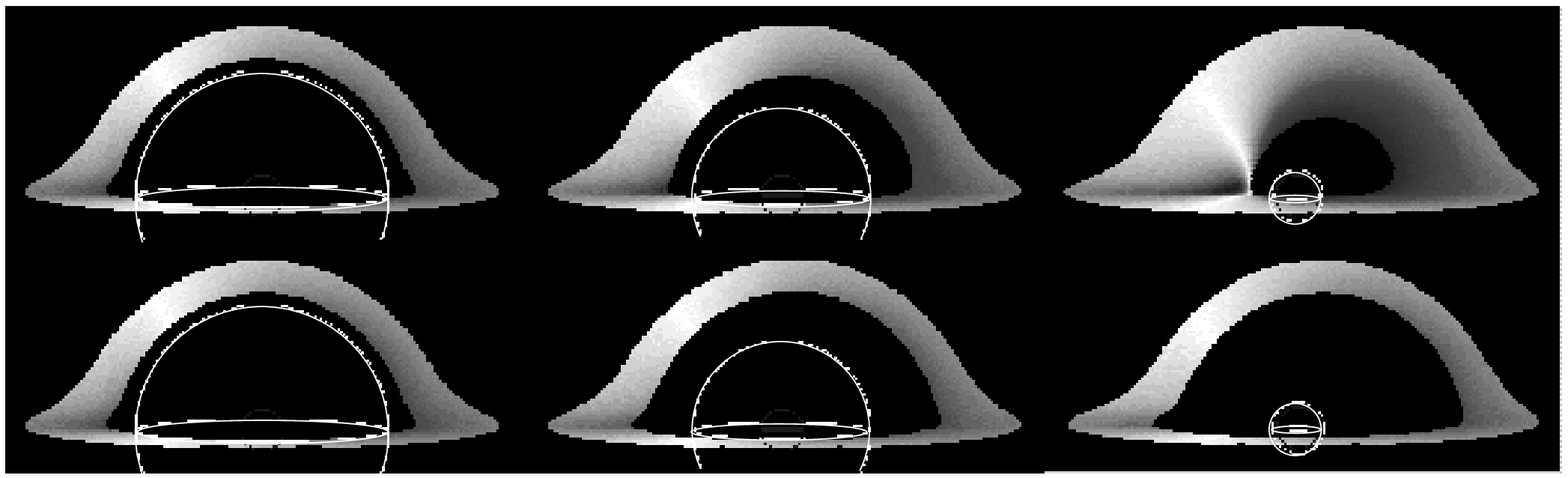}
        \end{minipage}
\sidebyside
{\caption{
Redshift maps for disks at inclination angle $85\deg$ with different
outer radii. 
Top to bottom: disks with outer radii $10R_{\mathrm g}$, $20R_{\mathrm g}$, 
$100R_{\mathrm g}$.
Left to right: $a=0.01, 0.5, 0.998$.
See text for details.
}
\label{fig:overallshape}}
{\caption{
Redshift maps for 10 $R_{\mathrm g}$  disks with different inner radii.
Top row: inner radii are the radii of the last stable orbits;
Bottom row: inner radii are 6 $R_{\mathrm g}$.
Left to right: $a=0.01, 0.5, 0.998$.
}
\label{fig:inner_contour}}
\vskip -0.1in
\end{figure}

Figures~\ref{fig:redshiftmap}, \ref{fig:overallshape} and 
\ref{fig:inner_contour} are all redshift maps.
Figures~\ref{fig:redshiftmap} shows the redshift maps of disks with 
an outer radius $20R_{\mathrm g}$ ($R_{\mathrm g} \equiv \frac{GM}{c^2}$)
and an inner radius as the last stable orbit. 
For a small viewing angle, there is only significant (gravitational) 
red-shift in the inner region;
whereas for large viewing angles, Doppler shift becomes important,
significant blue- and red-shift can be seen on the whole disk. 
(It would be clearer in color figures.)
Also, for large viewing angles, far end of the disk bends toward the observer.

\vskip -0.1in
\vskip -0.1in

Figure~\ref{fig:overallshape} shows the overall shape of disks due to
Doppler and gravitational effects at a large inclination angle ($85\deg$). 
The inner radius takes the 
value of the last stable circular orbit. In the first row, 
the white circles and ellipses show the last stable orbit.
The inner contours are greatly distorted by the focusing effect.

Figure~\ref{fig:inner_contour} shows the comparison between 
Shwarzschild black holes and Kerr black holes, which shows clearly
additional distortion due 
to frame-dragging effect. Spin results in a smaller last stable orbit,
and the inner contour changes from symmetric for $a=0.01$ to
asymmetric for $a=0.998$.
Therefore in principle, we can tell a rotating black hole from a 
non-rotating one from the inner contours of their accrection disks.

\vskip -0.1in

\section{Images of thin accretion disks}
\label{section:images}

With the redshift map, we calculated images of accretion disks 
around black holes in several energy ranges. In the calculations, 
we assume 1) the disks are standard thin disks (Shakura \& Sunyaev, 1973);
and 2) local radiation results from gravitational potential energy release
through blackbody radiation (Page \& Thorne 1974). 
The inner disk radii are the respective last stable orbits.

\vskip -0.1in
\begin{figure}[ht]
	\centering
        \begin{minipage}[t]{.48\textwidth}
                \centering
                \includegraphics[width=\textwidth]{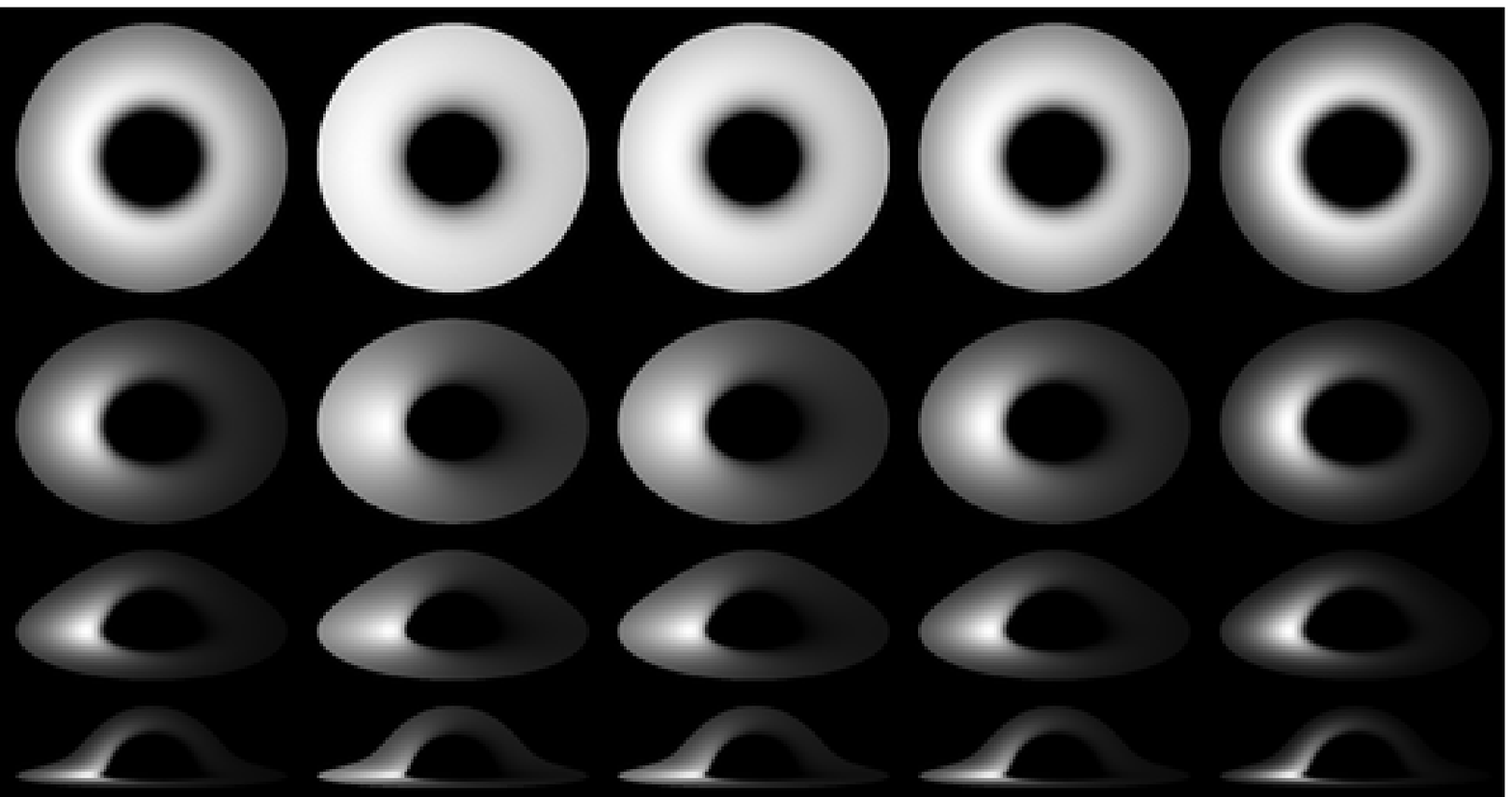}
a) $a=0.01$
\vskip 0.1in
                \includegraphics[width=\textwidth]{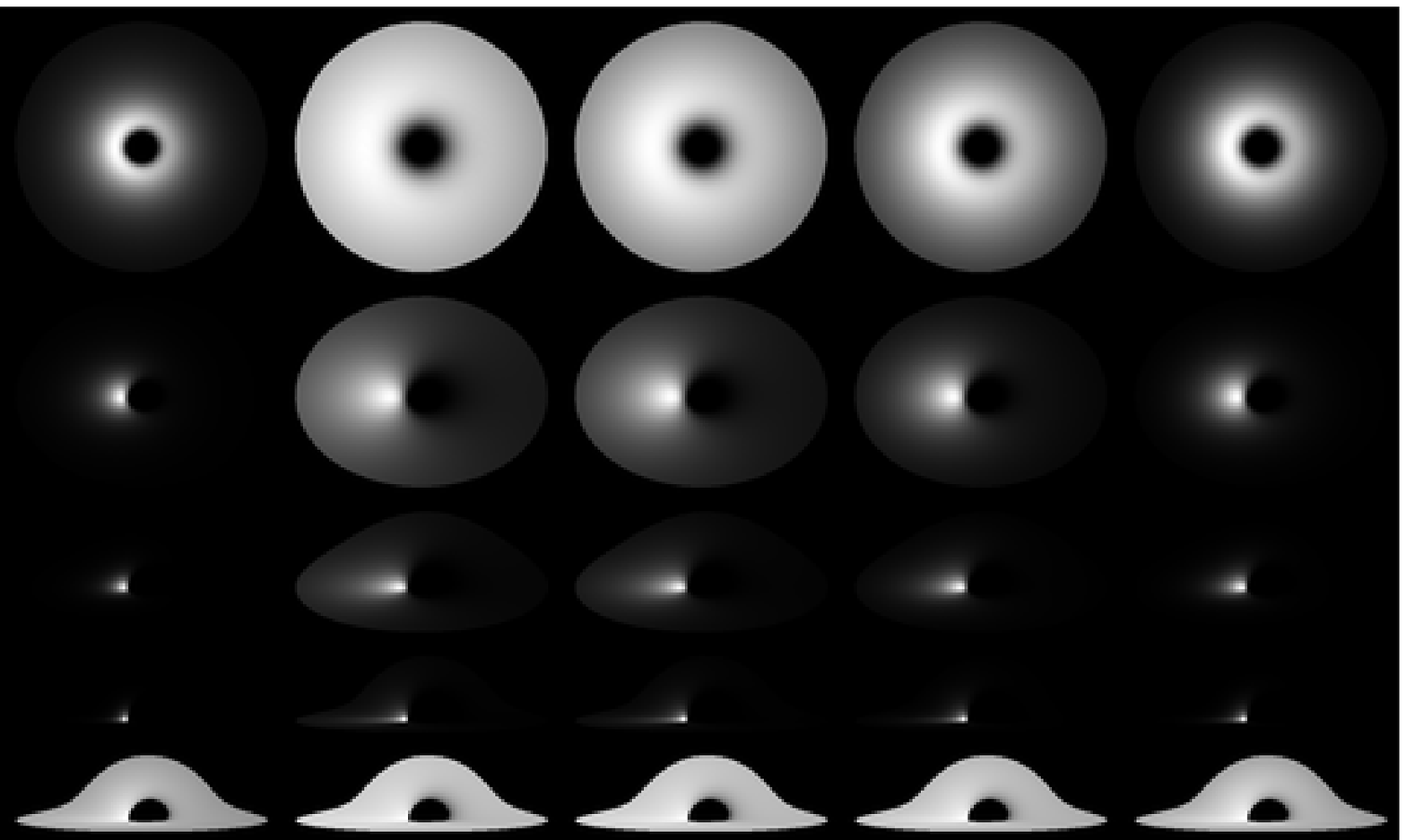}
                \includegraphics[width=\textwidth]{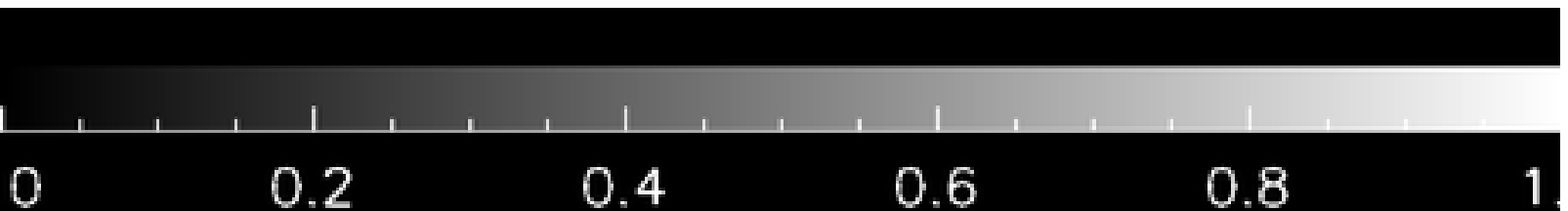}
c) $a=0.998$
        \end{minipage}
        \begin{minipage}[t]{.48\textwidth}
                \centering
                \includegraphics[width=\textwidth]{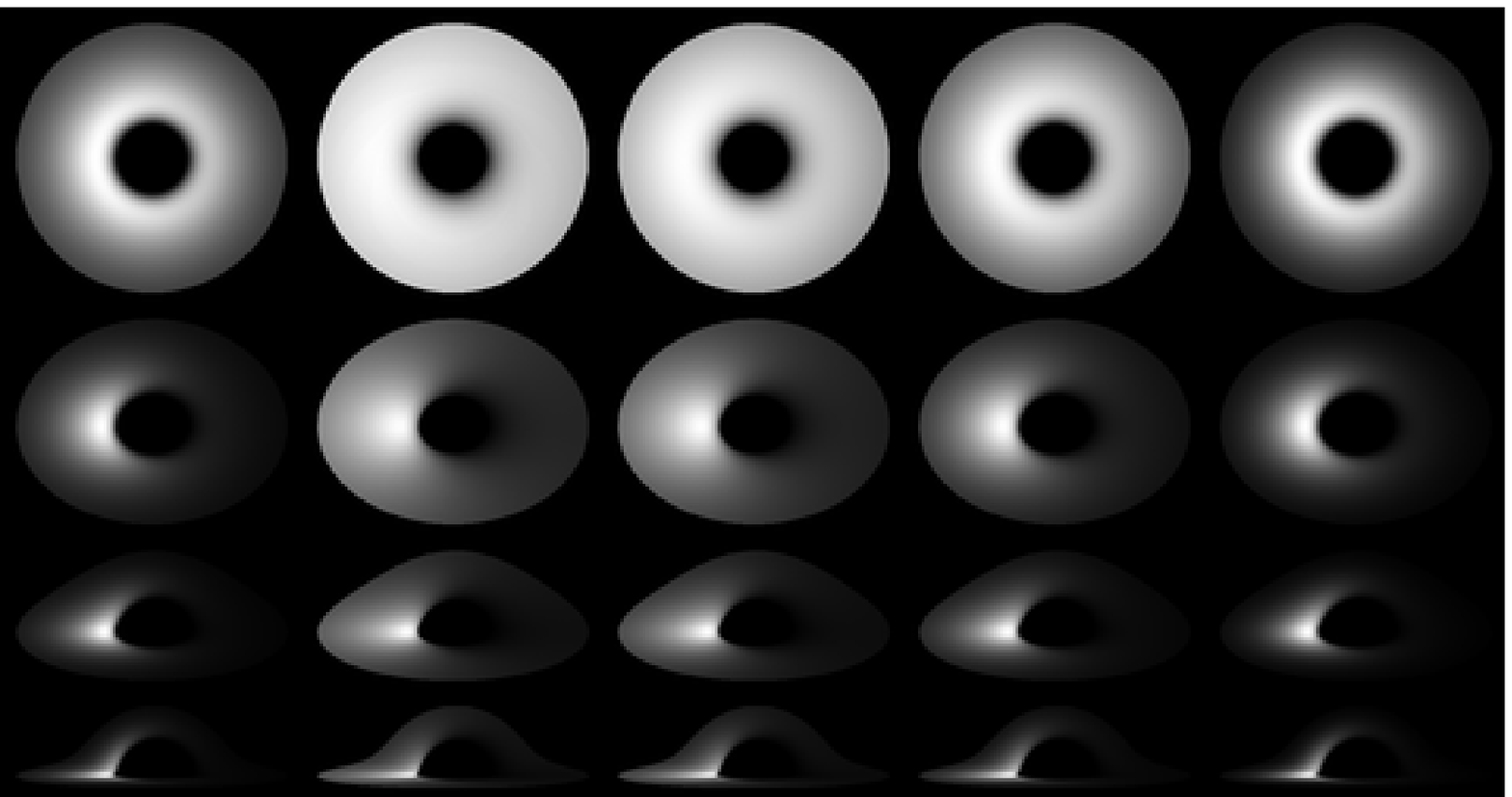}
b) $a=0.5$
\caption{Disk images of $20R_{\mathrm g}$ disks for different spin
of the black holes, viewing in certain energy ranges. In all figures,
same energy range in a column, and same inclination angle in a row.
Left to right: 0.01--100, 0.1--1, 1--2, 2--5, 5--10.
Top to bottom: $\theta=5\deg, 45\deg, 70\deg, 85\deg, 85\deg$. 
The last image row of c) is the same as the row above it, expect that
it is in log scale. c) also shows the scale bar.}
\label{figure:r20_images}
        \end{minipage}
\end{figure}
\vskip -0.1in

In all the images, energy ranges are in the unit of the energy 
corresponding to the maximum 
temperature in a non-rotating black hole case. For example, if the maximum 
temperature in a non-rotating black hole case is 1 keV, then the above 
energy ranges are 0.1 to 1 keV, 1 to 2 keV etc.. The maximum temperature 
in the extremal Kerr black hole case is about 5 keV.
The images show the brightness in gray scale,
with energy flux linearly 
(if not otherwise specified) mapped to the [0,1] shown on the scale bar
in Figure~\ref{figure:r20_images}c. 

\begin{figure}[ht]
	\centering
        \begin{minipage}{.48\textwidth}
                \centering
                \includegraphics[width=\textwidth]{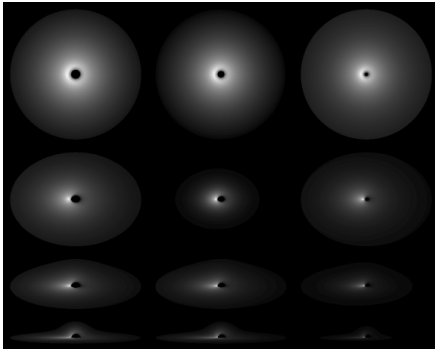}
        \end{minipage}
        \begin{minipage}{.48\textwidth}
                \centering
\caption{Disk images for $100R_{\mathrm g}$ disks in the energy range
0.5 -- 2. 
Top to bottom: $\theta=5\deg, 45\deg, 70\deg, 85\deg$.
Left to right: $a=0.01, 0.5, 0.998$ respectively.
0.5 to 2 keV is the energy range of NASA's
next generation mission MAXIM (Micro-Arcsecond X-ray Imaging Mission) 
pathfinder.
(The images in this figure have a lower contrast than the other images.)
}
        \end{minipage}
\label{fig:100Rg}
\vskip -0.2in
\end{figure}

\section{Summary and discussion}
\label{section:summary}
We studied the images of standard thin accretion disks around 
black holes under the strong relativistic effects. 
The expected images for systems with different black hole spin, 
viewing from different viewing angles 
and in different energy bands are given.
The spin of the black hole not only determines how close the disk can extend
to the center, but also causes additional distortion due to the dragging
of the frame.

The secondary image are left out because in most cases it will be blocked
by the disk, though it might be important in some cases.

\noindent{\bf Acknowledgments}

The authors would like to thank Dr. Fanton for providing us his code of calculating the redshift map and line profile of a thin disk. 
This study is supported in part by NASA's Marshall Space Flight Center 
and through NASA's Long Term Space Astrophysics Program.

{
\bibliographystyle{apalike}
\chapbblname{chapbib}
\chapbibliography{logic}
}

\end{document}